\documentclass[conference]{IEEEtran}

\usepackage{graphicx,psfrag,epsfig,epsf,latexsym,hhline,amsmath,amssymb,multirow}
\usepackage{pst-plot}
\usepackage{pstricks-add}
\usepackage{pifont}
\usepackage{graphicx}
\usepackage{amsthm}
\usepackage[linesnumbered,ruled,vlined]{algorithm2e}
\usepackage[noadjust]{cite}
\usepackage{array}
\usepackage{blindtext}
\usepackage{etoolbox}
\usepackage{comment}
\usepackage{epstopdf}

\SetCommentSty{mycommfont}

\SetKwInput{KwInput}{Input}                
\SetKwInput{KwOutput}{Output}              

\newcolumntype{M}[1]{>{\centering\arraybackslash}m{#1}}
\newcolumntype{P}[1]{>{\centering\arraybackslash}p{#1}}

\newcommand{\msf}[1]{\mathsf{#1}}

\newcommand{\SNR}{\text{SNR}}

\newcommand{\E}{\mathbb{E}}

\newcommand{\iid}{i.\@i.\@d.\ }

\theoremstyle{definition}
\newtheorem{lemma}{Lemma}

\newtheorem{theorem}{Theorem}
\newtheorem{remark}{Remark}

\newcommand\xqed[1]{%
  \leavevmode\unskip\penalty9999 \hbox{}\nobreak\hfill
  \quad\hbox{#1}}
\newcommand\demo{\xqed{$\blacksquare$}}

\IEEEoverridecommandlockouts
\begin{document}
\title{Coexistence of Heterogeneous Services in the Uplink with Discrete Signaling and Treating Interference as Noise}
\author{ Min Qiu$^\dag$, Yu-Chih Huang$^*$, and Jinhong Yuan$^\dag$\\
$^\dag$School of Electrical Engineering and Telecommunications, University of New South Wales, Sydney, Australia\\
$^*$Institute of Communications Engineering, National Yang Ming Chiao Tung University, Hsinchu City, Taiwan\\
E-mail: \{min.qiu, j.yuan\}@unsw.edu.au, jerryhuang@nycu.edu.tw  \\

\thanks{The work of Min Qiu and Jinhong Yuan was supported in part by the Australian Research Council (ARC) Discovery Project under Grant DP220103596, and in part by the ARC Linkage Project under Grant LP200301482. The work of Yu-Chih Huang was supported by the National Science and Technology Council, Taiwan, under Grant 111-2221-E-A49-069-MY3. This work was also supported in part by the Higher Education Sprout Project of the National Yang Ming Chiao Tung University and Ministry of Education (MOE), Taiwan.}

}

\maketitle

\begin{abstract}
The problem of enabling the coexistence of heterogeneous services, e.g., different ultra-reliable low-latency communications (URLLC) services and/or enhanced mobile broadband (eMBB) services, in the uplink is studied. Each service has its own error probability and blocklength constraints and the longer transmission block suffers from heterogeneous interference. Due to the latency concern, the decoding of URLLC messages cannot leverage successive interference cancellation (SIC) and should always be performed before the decoding of eMBB messages. This can significantly degrade the achievable rates of URLLC users when the interference from other users is strong. To overcome this issue, we propose a new transmission scheme based on discrete signaling and treating interference as noise decoding, i.e., without SIC. Guided by the deterministic model, we provide a systematic way to construct discrete signaling for handling heterogeneous interference effectively. We demonstrate theoretically and numerically that the proposed scheme can perform close to the benchmark scheme based on capacity-achieving Gaussian signaling with the assumption of perfect SIC.
\end{abstract}


%

\section{Introduction}\label{sec:intro}
The fifth generation (5G) wireless communication systems and beyond are expected to support a variety of service types such as enhanced mobile broadband (eMBB) and ultra-reliable low-latency communications (URLLC). To this end, the idea of enabling the coexistence of heterogeneous services in the same radio access network has attracted many interests \cite{8004168,8476595,8647460,9097306,9562192,9838392}. The current proposal is to slice the network by allocating orthogonal resources, e.g., in time/frequency domain, for each service \cite{8004168}. Since different services are isolated from each other, their own quality-of-service requirements are guaranteed. However, this approach could lead to very low spectral and energy efficiency when the number of devices is large.

Various research has been carried out to investigate efficient coexistence mechanisms for heterogeneous services with improved efficiency and user fairness. Notably, \cite{8476595,8647460} introduced a non-orthogonal uplink transmission scheme for enabling simultaneous communication between different types of services and the base station. In the Gaussian multiple access channel (MAC) with homogeneous and infinite blocklength, the entire capacity region can be efficiently achieved by superposition coding and successive interference cancellation (SIC), together with time-sharing \cite{Cover:2006:EIT:1146355}. However, as pointed out in \cite{8476595,8647460}, one salient difference between the conventional and heterogeneous cases is that the decoding of a URLLC transmission block cannot leverage SIC due to the latency constraints. That is, the base station should always decode URLLC services first before decoding eMBB of services regardless of channel conditions. Without SIC, the achievable rates of URLLC users could be severally degraded when the interference is strong. Note also that SIC can fail under finite blocklength constraint \cite{5452208,8345745}. In addition, to ensure ultra reliability for URLLC services, performing SIC by decoding eMBB first then URLLC would require the decoding of eMBB services to achieve at least the same reliability as for URLLC services. However, this is an overkill as eMBB messages are coded based on a lower reliability requirement than URLLC services. Further, SIC can introduce extra decoding delay, complexity, and error propagation, which can become pronounced when the number of users is large. In light of the above, the non-orthogonal schemes relying on SIC may not fully address the challenges from the coexistence of URLLC and other services; thereby, new proposals are called for.

The achievable rate region of the Gaussian MAC under finite blocklength in terms of the second-order term and the third-order term were rigorously derived in \cite{7300429} and \cite{9535162}, respectively. In order to achieve the optimal first- and second-order rates, both works considered the input distribution to be shell codes, i.e., codewords drawn from a power shell. Moreover, the achievability bounds therein were derived under joint typically and joint maximum-likelihood decoding, respectively, with global error probability formalism and homogeneous blocklength constraints. Nevertheless, these results are not applicable to the case with heterogeneous error probability and blocklength constraints. Moreover, joint decoding has a much higher complexity than SIC decoding, which is difficult to realize in practice.

Motivated by the benefits and challenges of heterogeneous services coexistence in the uplink, we aim to design a new and practical scheme based on discrete signaling and treating interference as noise (TIN) decoding. We emphasize that the proposed discrete signaling is formed by practical channel coding and discrete constellations, e.g., quadrature amplitude modulation (QAM), which are the prevailing setup in current communication systems \cite{TS138212_v16p8}. Meanwhile, the TIN decoding only has single-user decoding complexity and latency, which is more suitable for decoding URLLC services. The designs of discrete signaling with TIN with infinite blocklength and homogeneous interference were investigated for the Gaussian broadcast channel (GBC) and the Gaussian interference channel with infinite blocklength in \cite{8291591} and \cite{9535131}, respectively. Recently, we proposed a transmission scheme for the GBC under heterogeneous blocklength and error probability constraints \cite{JSACQiu22}. However, the scheme therein cannot be applied here due to the subtle difference between the broadcast nature (i.e., all users' signals go through the same channel to one receiver) and the multiple access nature (i.e., each user's signal goes through a different channel to the receiver). Hence, it remains unclear on how to design the discrete signaling for each uplink user to effectively handle heterogeneous interference with TIN to achieve a rate close to that under Gaussian signaling with perfect SIC.

In this paper, we provide the design and analysis of a new multiple access scheme based on discrete signaling and TIN decoding to support heterogeneous services coexistence in the uplink. We first approximate the MAC channel as a linear deterministic model \cite{Avestimehr11} and design a capacity-achieving coding scheme with TIN. We then translate the scheme from the deterministic model into the coded modulation scheme with TIN for the MAC channel model under heterogeneous constraints. The main feature of the proposed scheme is that the user with a longer transmission block can use different sets of constellations in different parts of its block depending on heterogeneous interference statistics. Whereas the encoding and decoding are single-user based. Our theoretical and numerical results show that the proposed scheme can achieve rate pairs close to the benchmark scheme that assumes Gaussian signaling and perfect SIC.

\subsection{Notations}
Random variables are written in upper case san-serif fonts, e.g., $\msf{X}$, while their realizations are in lower case form, e.g., $x$. $\msf{X}_1 \overset{d}{=}\msf{X}_2$ means that $\msf{X}_1$ and $\msf{X}_2$ are equal in distribution. All logarithms are base 2. $\msf{X}\overset{\text{unif}}{\sim} \text{QAM}(|\Lambda|,d_{\min}(\Lambda))$ represents that $\msf{X}$ is uniformly distributed over a zero mean regular QAM $\Lambda$ with cardinality $|\Lambda|$, minimum distance $d_{\min}(\Lambda)$, and average energy $E_{\Lambda} = d_{\min}^2(\Lambda)\frac{|\Lambda|-1}{6}$. $\lceil x\rceil$ rounds $x$ to the
nearest integer greater than or equal to $x$. We define the operation $(x)^+ \triangleq \max\{0,x\}$. The binary field, the collections of binary vectors of size $n$ and binary matrices of size $m\times n$ are denoted by $\mathbb{F}_2$, $\mathbb{F}^n_2$, and $\mathbb{F}_2^{m,n}$, respectively.


\section{System Model}\label{sec:model}
We consider a two-user uplink MAC that consists of two senders and one receiver. We leave the extension of our work to the $K$-user case in our future work. Let $k\in \{1,2\}$ be the user index. Let $\boldsymbol{x}_k\in\mathbb{C}^{N_k}$ represent the coded symbols of user $k$, where $N_k$ is the symbol length. Due to heterogeneous blocklength constraints, we assume $N_1 \leq N_2$ without loss of generality. Each user's coded symbols satisfy the individual maximum power constraint per codeword given by
\begin{align}\label{eq:p_constraint}
\frac{1}{N_k}\sum^{N_k}_{j=1}|x_k[j]|^2\leq P_k.
\end{align}

Let $h_k\in\mathbb{C}$ represent the complex channel for user $k$. The $j$-th received symbol is
\begin{align}\label{eq:model}
y[j]= \left\{ \begin{array}{l}
h_1x_1[j]+h_2x_2[j]+z[j],\;j=1,\ldots,N_1\\
h_{2}x_{2}[j]+z[j],j=N_1+1,\;\ldots,N_2\\
\end{array} \right.,
\end{align}
where $z[j] \sim \mathcal{CN}(0,1)$ is the i.i.d. Gaussian noise for $j\in \{1,\ldots,N_{2}\}$. In addition, we assume the channel state information is available so that each transmitter can eliminate the channel phase by rotating its signal by $\frac{h_k^*}{|h_k|}$. In this regard, the channel model in \eqref{eq:model} can be equivalently expressed as that with some $h_k\in \mathbb{R}$.

From \eqref{eq:model}, it can be seen that the first $N_1$ symbols of $\boldsymbol{x}_{2}$ and $\boldsymbol{x}_1$ are superimposed. This is because we consider that user 2, e.g., a URLLC user, needs to transmit its signals as soon as possible due to its urgency. In this case, the received symbol block of user 2 suffers from \emph{heterogeneous interference} across its symbols. Note that user 2 can be a different URLLC user or a user of other service types that are less urgent, e.g., eMBB. We denote by $\SNR_k=P_k|h_k|^2$, $R_k$, and $\epsilon_k$ the signal-to-noise ratio (SNR), achievable rate, and the requirement of the decoding error probability of user $k$, respectively.



\section{Proposed Discrete Signaling Scheme with TIN}\label{sec:scheme}
In this section, we first present the finite blocklength achievable rate of the MAC with an arbitrary choice of discrete input distributions, blocklength, and error probability requirements. Using the insight obtained from the finite blocklength analysis, we approximate the considered channel model by using the linear deterministic model \cite{Avestimehr11} and construct a capacity-achieving coding scheme with TIN. Next, we translate the proposed scheme of the deterministic channel to the coding scheme for the original MAC under heterogeneous constraints. We show analytically that the gap between the mutual information under TIN and the corresponding single-user capacity is upper bounded by a constant gap.

\subsection{Finite Blocklength Achievable Rate}\label{sec:FBL}
In this work, we consider that both users employ discrete and finite constellations. Specifically, user 1's symbol satisfies $\msf{X}_1[j]\overset{\text{unif}}{\sim}\Lambda_1$ for $j=1,\ldots,N_1$, where $\Lambda_1$ is user '1s constellation set. In contrast, user 2 uses two sets of constellations to handle heterogeneous interference. Thus, its symbol satisfies $\msf{X}_{2}[j]\overset{\text{unif}}{\sim}\Lambda_{2,1}$ for $j=1,\ldots,N_1$ and $\msf{X}_{2}[j]\overset{\text{unif}}{\sim}\Lambda_{2,2}$ for $j=N_1+1,\ldots,N_{2}$. For ease of presentation, we let $\msf{X}_{2,1}\overset{d}{=}X_{2}[j]$ and $\msf{Y}_1\overset{d}{=} \msf{Y}[j]$ for $j=1,\ldots,N_1$, and let $\msf{X}_{2,2}\overset{d}{=} X_{2}[j]$ and $\msf{Y}_2\overset{d}{=} \msf{Y}[j]$ for $j=N_1+1,\ldots,N_{2}$.

The finite blocklength achievable rate pairs given constellation tuples $(\Lambda_1,\Lambda_{2,1},\Lambda_{2,2})$ are provided in Theorem \ref{prof:u1}.
\begin{theorem}\label{prof:u1}
Let $\epsilon_k$ be the upper bound on the average TIN decoding error probability of user $k$. For the channel model in \eqref{eq:model}, the achievable rates of users 1 and 2 with TIN are
\begin{align}
R_1 \leq& I(\msf{X}_1;\msf{Y}_1)
-\sqrt{\frac{V(\msf{X}_1;\msf{Y}_1)}{N_1}}Q^{-1}(\epsilon_1)+O\left(\frac{\log N_1}{N_1}\right),\label{eq:ukrate} \\
R_{2} \leq& \frac{N_1 I(\msf{X}_{2,1};\msf{Y}_1)+(N_{2}-N_1)I(\msf{X}_{2,2};\msf{Y}_2) }{N_{2}} \nonumber\\
&-\frac{\sqrt{N_1V(\msf{X}_{2,1};\msf{Y}_1)+(N_{2}-N_1)V(\msf{X}_{2,2};\msf{Y}_2)}}{N_{2}}\nonumber\\
&\times Q^{-1}\left(\epsilon_{2}\right)+O\left(\frac{\log N_{2}}{N_{2}}\right), \label{eq:u1rate}
\end{align}
where we have user 2's mutual information $I(\msf{X}_{2,t};\msf{Y}_t)=\E[i(\msf{X}_{2,t};\msf{Y}_t)]$ and dispersion $V(\msf{X}_{2,t};\msf{Y}_t)=\text{Var}[i(\msf{X}_{2,t};\msf{Y}_t)]$ for $t\in\{1,2\}$ with information densities $i(\msf{X}_{2,1};\msf{Y}_1)=\log \left( \frac{\sum_{x_{2,1}\in\Lambda_{2,1}}e^{-|y-h_1x_1-h_{2}x_{2,1}|}}{\frac{1}{|\Lambda_{2,1}|}\sum_{x_{2,1}\in\Lambda_{2,1}}\sum_{x_1\in\Lambda_1}e^{-|y-h_1x_1-h_{2}x_{2,1}|}}\right)$, and $i(\msf{X}_{2,2};\msf{Y}_2)=\log \left(\frac{e^{-|y-h_{2}x_{2,2}|}}{\frac{1}{|\Lambda_{2,2}|}\sum_{x_{2,2}\in\Lambda_{2,2}}e^{-|y-h_{2}x_{2,2}|}}\right)$, and $Q^{-1}(x)$ is the inverse of Q function $Q(x)=\int^{\infty}_{x}\frac{1}{\sqrt{2\pi}}e^{-\frac{t^2}{2}}dt$. User 1's mutual information $I(\msf{X}_1,\msf{Y}_1)$ and dispersion $V(\msf{X}_1;\msf{Y}_1)$ can be easily obtained from $I(\msf{X}_{2,1};\msf{Y}_1)$ and $V(\msf{X}_{2,1};\msf{Y}_1)$ by swapping the argument between subscripts ``1'' and ``2''.
\end{theorem}
The proof of Theorem \ref{prof:u1} follows from the steps in \cite[Sec. V]{JSACQiu22} and is omitted due to space limitation. 

\begin{remark}
The impacts of the interfering symbol blocklength on the achievable rate is clearly shown in \eqref{eq:u1rate}. Under heterogeneous interference, the achievable rate is no longer determined by a single signal-to-interference-plus-noise ratio (SINR) as in the homogeneous blocklength case. In addition, the relationship between heterogeneous blocklengths and achievable rate in \eqref{eq:u1rate} holds for other i.i.d. inputs, e.g., i.i.d. Gaussian codes. However, for the shell codes commonly used for deriving larger finite blocklength achievable rates in many works \cite{5452208,7300429,9535162}, this relation may no longer hold as the distribution of each symbol of shell codes is not independent.
\demo
\end{remark}

From Theorem \ref{prof:u1}, we note that for given power, blocklength, and error probability constraints, larger mutual information and smaller dispersion are desirable for achieving a larger rate. In the subsequent sections, we introduce the proposed design and justify its effects on mutual information and dispersion.

\subsection{Deterministic MAC under Heterogeneous Interference}\label{sec:det_model}
Theorem \ref{prof:u1} hints that to attain a larger achievable rate, different signaling designs are required for interfering and interference-free parts. To this end, we approximate the MAC channel under heterogeneous blocklength constraints as a concatenation of two deterministic channels and propose a capacity-achieving coding scheme with TIN. Let $n_k \triangleq \lceil \log  \SNR_k \rceil^+$ be the approximated single-user capacity of users $k$, $k\in\{1,2\}$. By adopting the idea in \cite{Avestimehr11}, the channel model of \eqref{eq:model} is approximated as
\begin{align}
\msf{Y}_1=& \boldsymbol{S}^{q-n_1}\msf{X}_1 \oplus \boldsymbol{S}^{q-n_{2}}\msf{X}_{2,1}, \label{eq:det_model1}\\
\msf{Y}_2=&\msf{X}_{2,2},\label{eq:det_model2}
\end{align}
where the multiplication and summation $\oplus$ are over $\mathbb{F}_2$, $\msf{X}_1,\msf{X}_{2,1},\msf{Y}_1 \in \mathbb{F}_2^q$, $ \msf{X}_{2,2},\msf{Y}_2 \in \mathbb{F}_2^{n_{2}}$ are binary column vectors and the subscripts ``1'' and ``2'' follow those defined in the first paragraph of Sec. \ref{sec:FBL}, $q=\max\{n_1,n_{2}\}$, and $\boldsymbol{S}$ is a $q \times q$ down shift matrix. The operational meaning of the binary column vector is that each of its entries represents a power level or a bit. $\boldsymbol{S}^{q-n_1}\msf{X}_1$ models the channel effect, where the lowest $q-n_1$ bits of $\msf{X}_1$ are shifted down below the noise level and truncated.

We first assume $|h_1|>|h_{2}|$ such that $q=n_1$. For the channel model in \eqref{eq:det_model1} and \eqref{eq:det_model2}, the set of non-negative rate tuples $(m_1,m_{2,1},m_{2,2})$ are achievable if
\begin{align}
m_1+m_{2,1} \leq& n_1, \label{eq:cap1}\\
m_{2,1} \leq& n_{2},\label{eq:cap2}\\
m_{2,2} \leq & n_{2}.\label{eq:cap3}
\end{align}
Observe that \eqref{eq:cap1} and \eqref{eq:cap2} are the deterministic MAC capacity region and \eqref{eq:cap3} is the deterministic point-to-point channel capacity, which are within 1 bit of the capacity of their Gaussian counterparts \cite{Avestimehr11}, respectively. It is worth emphasizing that the considered deterministic MAC is different from the one in \cite{Avestimehr11} where all user 2's bits experience the same level of interference strength. Different from the achievable scheme that relies on SIC \cite{Avestimehr11}, we show how to design input distributions $(\msf{X}_1,\msf{X}_{2,1},\msf{X}_{2,2})$ to achieve the above rate region with TIN.

Let $\msf{U}_1\in\mathbb{F}^{m_1}_2$ and $\msf{U}_{2,1}\in\mathbb{F}^{m_{2,1}}_2$ be users 1 and 2's message vectors, respectively, with each entry drawn independently and uniformly distributed from $\mathbb{F}_2$. Let $\boldsymbol{G}_1 \in \mathbb{F}_2^{q,m_1}$ and $\boldsymbol{G}_{2,1} \in \mathbb{F}_2^{q,m_{2,1}}$ be the generator matrices such that $\msf{X}_1 = \boldsymbol{G}_1 \msf{U}_1$ and $\msf{X}_{2,1} = \boldsymbol{G}_{2,1} \msf{U}_{2,1}$. The achievable rate of user 1 under TIN is
\begin{align}\label{eq:I1}
&I(\msf{X}_1;\msf{Y}_1) = H(\msf{Y}_1) - H(\msf{Y}_1|\msf{X}_1) \nonumber \\
 =& H(\boldsymbol{S}^{q-n_1}\boldsymbol{G}_1 \msf{U}_1\oplus \boldsymbol{S}^{q-n_{2}}\boldsymbol{G}_{2,1} \msf{U}_{2,1})
 - H(\boldsymbol{S}^{q-n_{2}}\boldsymbol{G}_{2,1} \msf{U}_{2,1}) \nonumber  \\
 =& \text{rank}([\boldsymbol{S}^{q-n_{k}}\boldsymbol{G}_1,\boldsymbol{S}^{q-n_{2}}\boldsymbol{G}_{2,1}])
 - \text{rank}(\boldsymbol{S}^{q-n_{2}}\boldsymbol{G}_{2,1}),
\end{align}
where the multiplication and addition are over $\mathbb{F}_2$. User 2's rate under TIN $I(\msf{X}_{2,1};\msf{Y}_1)$ can be easily obtained from \eqref{eq:I1} by swapping the arguments between subscripts ``1'' and ``2''.

To achieve the rate region of \eqref{eq:cap1}-\eqref{eq:cap2}, we propose
\begin{align}\label{eq:G1G2}
\boldsymbol{G}_1\hspace{-1mm}=\hspace{-1mm}\begin{bmatrix}
\boldsymbol{F}_{1,1}\\
\boldsymbol{0}^{n_{1}-n_{2}-m_{1,1},m_1}\\
\boldsymbol{0}^{m_{2,1},m_1}\\
\boldsymbol{0}^{n_{2}-m_{2,1}-m_{1,2},m_1}\\
\boldsymbol{F}_{1,2}
\end{bmatrix},
\boldsymbol{G}_{2,1}\hspace{-1mm}=\hspace{-1mm}\begin{bmatrix}
\boldsymbol{F}_{2,1}\\
\boldsymbol{0}^{n_{2}-m_{2,1},m_{2,1}}\\
\boldsymbol{0}^{n_1-n_{2},m_{2,1}}\\
\end{bmatrix},
\end{align}
where $\boldsymbol{F}_{1,1}\in\mathbb{F}^{m_{1,1},m_1}_2$, $\boldsymbol{F}_{1,2}\in\mathbb{F}^{m_{1,2},m_1}_2$, $\boldsymbol{F}_{2,1}\in\mathbb{F}^{m_{2,1},m_{2,1}}_2$, are submatrices with linearly independent rows,  $m_{1,1}=\min\{(m_1+m_{2,1}-n_{2})^+,n_1-n_{2}\}$, $m_{1,2}=\min\{m_1,n_{2}-m_{2,1} \}$ such that $m_1 = m_{1,1}+m_{1,2}$. Substituting the proposed $\boldsymbol{G}_1$ and $\boldsymbol{G}_{2,1}$ into \eqref{eq:I1}, we have that
\begin{align}
\text{rank}&([\boldsymbol{S}^{q-n_{1}}\boldsymbol{G}_1,\boldsymbol{S}^{q-n_{2}}\boldsymbol{G}_{2,1}])  \nonumber \\
&=\text{rank}\left(
\setlength\arraycolsep{0pt}
\begin{bmatrix}
\left. {\begin{array}{*{20}{c}}
\boldsymbol{F}_{1,1}\\
\boldsymbol{0}^{n_{k}-n_{2,1}-m_{1,1},m_1}
\end{array}} \right\} & \boldsymbol{0}^{n_1-n_{2},m_{2,1}} \\
\boldsymbol{0}^{m_{2,1},m_1} & \boldsymbol{F}_{2,1} \\
\left. {\begin{array}{*{20}{c}}
\boldsymbol{0}^{n_{2}-m_{2,1}-m_{1,2},m_1}\\
\boldsymbol{F}_{1,2}
\end{array}} \right\} & \boldsymbol{0}^{n_{2}-m_{2,1},m_{2,1}}\\
\end{bmatrix}
\right)\\
&= m_1+m_{2,1}, \label{eq:1_term}\\
\text{rank}&(\boldsymbol{S}^{q-n_{2}}\boldsymbol{G}_{2,1})=\text{rank}(\boldsymbol{F}_{2,1}) =m_{2,1}. \label{eq:2_term}
\end{align}
Substituting \eqref{eq:1_term} and \eqref{eq:2_term} into \eqref{eq:I1} results in $I(\msf{X}_1;\msf{Y}_1)=m_1$. Similarly, we can follow the above steps to obtain that $I(\msf{X}_{2,1};\msf{Y}_1)=m_{2,1}$.

Let $\msf{U}_{2,2}\in\mathbb{F}^{m_{2,2}}_2$ be another part of user 2's message vector. To achieve the rate in \eqref{eq:cap3}, we design
\begin{align}\label{eq:G22}
\boldsymbol{G}_{2,2}=\begin{bmatrix}
\boldsymbol{F}_{2,2}\\
\boldsymbol{0}^{n_{2}-m_{2,2},m_{2,2}}\\
\end{bmatrix},
\end{align}
where $\boldsymbol{F}_{2,2}\in\mathbb{F}^{m_{2,2},m_{2,2}}_2$. Then, we have that $\msf{X}_{2,2} = \boldsymbol{G}_{2,2} \msf{U}_{2,2}$, which is interference-free according to \eqref{eq:det_model2}. As a result, we have that $I(\msf{X}_{2,2};\msf{Y}_2)=\text{rank}(\boldsymbol{F}_{2,2})=m_{2,2}$. From here, we see that the proposed scheme can achieve the rate region of \eqref{eq:cap1}-\eqref{eq:cap3} with TIN.


In the subsequent sections, we focus on achieving the boundary points of this capacity region such that only the equalities of \eqref{eq:cap1} and \eqref{eq:cap3} are active. In this case, we set $m_{1,1} = n_1-n_{2}$, $m_{1,2}=n_{2}-m_{2,1}=m_1+n_{2}-n_1$, and $m_{2,2}=n_{2}$.



When $|h_1|<|h_{2}|$, the design of $\boldsymbol{G}_1$ and $\boldsymbol{G}_{2,1}$ can be easily obtained from \eqref{eq:G1G2} by swapping the argument between subscripts ``1'' and ``2''. The design of $\boldsymbol{G}_{2,2}$ remains unchanged.

The above deterministic approach allows us to obtain a systematic design on the input distributions for the original model, which we will show in the next section.

\subsection{Proposed Coded Modulation Scheme With TIN}
We translate the proposed scheme in Sec. \ref{sec:det_model} into the coded modulation scheme for the MAC under heterogeneous interference. We assume $|h_1|>|h_{2}|$ for illustrate purposes. The proposed scheme consists of the following steps.

\subsubsection{Encoding}
User $k$ encodes its message into codeword $\boldsymbol{c}_k$ by using a length-$L_k$ rate-$r_k$ binary code. Then, codeword $\boldsymbol{c}_k$ is interleaved by employing bit-interleaved coded modulation (BICM) technique \cite{CIT-019}. The interleaved codeword $\tilde{\boldsymbol{c}}_k$ is mapped to a length-$N_k$ symbol sequence $\boldsymbol{x}_k$, where the modulation design and mapping steps will be described in Sec. \ref{sec:2u_const} and Sec. \ref{sec:2u_const_map}, respectively. It is worth emphasizing that each user only uses a \emph{single} channel code such that the encoding complexity is the same as for the single-user case.

\subsubsection{Modulation Design}\label{sec:2u_const}
The proposed scheme for the deterministic model in Sec. \ref{sec:det_model} is systematically translated into the following QAM signaling
\begin{align}
\msf{X}_1[j] &= \eta \sqrt{P_1} (\msf{F}_{1,2}+2^{\frac{n_{2}}{2}}\msf{F}_{1,1}) \nonumber \\
&\overset{\text{unif}}{\sim}\Lambda_1,\;\text{for}\; j=1,\ldots,N_1,\label{eq:X1}\\
X_{2}[j] \hspace{-1mm}&=\hspace{-1mm}\left\{ \begin{array}{l}
\hspace{-2mm}\eta \sqrt{P_{2}}\cdot 2^{\frac{\log \SNR_1-m_{2,1}+n_{2}-\log \SNR_2}{2}}\msf{F}_{2,1}\hspace{-1mm}\overset{\text{unif}}{\sim}\hspace{-1mm}\Lambda_{2,1},\\
\quad\text{for}\;j=1,\ldots,N_1,\\
\hspace{-2mm}\eta'\sqrt{P_{2}} \msf{F}_{2,2}\overset{\text{unif}}{\sim}\Lambda_{2,2},\;\text{for}\;j\hspace{-1mm}=\hspace{-1mm}N_1\hspace{-1mm}+\hspace{-1mm}1,\;\ldots,N_{2},\\
\end{array}
\right.  \label{eq:X2}
\end{align}
where $\msf{F}_{k,t} \overset{\text{unif}}{\sim} \text{QAM}(2^{\text{rank}(\boldsymbol{F}_{k,t})},1)$ for $k,t\in\{1,2\}$ and the rank number follows \eqref{eq:G1G2} and \eqref{eq:G22}, $\Lambda_1$ is the superposition of two scaled regular QAM constellations with total cardinality $2^{m_1}$ whereas $\Lambda_{2,1}$ and $\Lambda_{2,2}$ are two scaled regular QAM constellations with cardinalities $2^{m_{2,1}}$ and $2^{m_{2,2}}$, respectively, $\eta$ and $\eta'$ are the normalization factors to ensure that $\E[| \msf{X}_k[j]|^2]\leq P_k$ for $j=1,\ldots,N_k$. Specifically, the normalization factors are
\begin{align}
\eta =& \sqrt{\frac{1}{\max\{E_{1}, E_{2,1}\}}}, \\
\eta'= &\frac{1}{\E\left[\left|\msf{F}_{2,2}\right|^2\right]}= \frac{6}{2^{n_{2}}-1},
\end{align}
where we define
\begin{align}
E_1\triangleq &\E\left[\left|\msf{F}_{1,2}+2^{\frac{n_{2}}{2}}\msf{F}_{1,1}\right|^2\right]\\
=&\frac{2^{n_{2}-m_{2,1}}-1+2^{n_1}-2^{n_{2}}}{6},\\
E_{2,1} \triangleq &\E\left[\left|2^{\frac{\log \SNR_1-m_{2,1}+n_{2}-\log \SNR_2}{2}}\msf{F}_{2,1}\right|^2\right]\\
 =&\frac{2^{n_{2}+\log \SNR_1-\log\SNR_2}(1-2^{-m_{2,1}})}{6}.
\end{align}
The power coefficient in front of each random variable $\msf{F}$ is obtained by counting the number of rows below its corresponding submatrix $\boldsymbol{F}$ in $\boldsymbol{G}$. For example, $2^{\frac{n_{2}}{2}}$ is due to that the number of rows below $\boldsymbol{F}_{1,1}$ in $\boldsymbol{G}_1$ is $n_{2}$. For $\msf{F}_{2,1}$, the number of rows below $\boldsymbol{F}_{2,1}$ in $\boldsymbol{G}_{2,1}$ is $n_1-m_{2,1} \approx \log \SNR_1-m_{2,1}+n_{2}-\log \SNR_2$. The reason for using these power coefficients will become clear in Sec. \ref{sec:analysis}.

For the case of $|h_1|<|h_{2}|$, the modulation design can be trivially obtained from \eqref{eq:X1} and \eqref{eq:X2} by swapping the argument between subscripts ``1'' and ``2''. 

\subsubsection{Modulation Mapping}\label{sec:2u_const_map}
From \eqref{eq:model}, we note that user 2's transmission block suffers from interference in the first $N_1$ symbols whereas the last $N_{2}-N_1$ symbols are interference-free. Thus, user 2 uses two sets of constellations as shown in \eqref{eq:X2} to handle heterogeneous interference. Following the modulation design in \eqref{eq:X2}, user 2 maps the first sub-block of the interleaved codeword $[\tilde{c}_{2}[1],\ldots,\tilde{c}_{2}[N_1m_{2,1}]]$ to $[x_{2}[1],\ldots,x_{2}[N_1]]$  and the last sub-block $[\tilde{c}_{2}[N_1m_{2,1}+1],\ldots,\tilde{c}_{2}[L_{2}]]$ to $[x_{2}[N_1+1],\ldots,x_{2}[N_{2}]]$, respectively. Likewise, user 1 maps $\boldsymbol{c}_1$ to $[x_1[1],\ldots,x_1[N_1]]$ following \eqref{eq:X1}.


\subsubsection{TIN Decoding}
At the receiver, each user's message is decoded by treating other user's signals as noise. Specifically, the receiver starts to decode user 1's message upon receiving $[y[1],\ldots,y[N_1]]$ while the decoding of user 2 starts upon receiving $[y[1],\ldots,y[N_{2}]]$. Unlike SIC decoding, TIN decoding can be performed in parallel, which is more favorable if both users are URLLC users. The decoding process is the same as that in the point-to-point channel using BICM \cite{CIT-019}.


\subsection{Performance Analysis}\label{sec:analysis}
In this section, we analyze the performance of the proposed scheme in the Gaussian MAC with heterogeneous blocklengths. We first bound the mutual information of user 1 with TIN decoding as follows.
\begin{align}
I(\msf{X}_1;\msf{Y}_1)=&h(\msf{Y}_1)-h(\msf{Y}_1|\msf{X}_1) \label{eq:I1_step1} \\
=&h(\msf{Y}_1)-h(Z) \nonumber \\
&-(h(h_{2}\msf{X}_{2,1}+Z)-h(Z)) \\
=&I(h_1\msf{X}_1+h_{2}\msf{X}_{2,1};\msf{Y}_1)\nonumber \\
&-I(h_{2}\msf{X}_{2,1};h_{2}\msf{X}_{2,1}+Z)\\
\geq&I(h_1\msf{X}_1+h_{2}\msf{X}_{2,1};\msf{Y}_1)-H(\msf{X}_{2,1}).\label{eq:I1_GMAC}
\end{align}
It remains to bound $I(h_1\msf{X}_1+h_{2}\msf{X}_{2,1};\msf{Y}_1)$. We note that $h_1\msf{X}_1+h_{2}\msf{X}_{2,1}$ is a two-dimensional discrete constellation, whose achievable rate can be bounded by applying \cite[Lemma 5]{9535131} as
\begin{align}\label{eq:I_dmin1}
I&(h_1\msf{X}_1+h_{2}\msf{X}_{2,1};\msf{Y}_1) \geq H(h_1\msf{X}_1+h_{2}\msf{X}_{2,1}) \nonumber\\
&-\log 2\pi e \left(\frac{1}{4}+\frac{4}{\pi d^2_{\min}(h_1\msf{X}_1+h_{2}\msf{X}_{2,1})}\right).
\end{align}
We can also easily obtain user 2's mutual information under TIN $I(\msf{X}_{2,1};\msf{Y}_1)$ from \eqref{eq:I1_GMAC} whose lower bound is also determined by $d_{\min}(h_1\msf{X}_1+h_{2}\msf{X}_{2,1})$ as in \eqref{eq:I_dmin1}. Thus, a large $d_{\min}(h_1\msf{X}_1+h_{2}\msf{X}_{2,1})$ is desirable for both users as it improves the mutual information lower bounds of both users under TIN decoding. With the proposed design in \eqref{eq:X1} and \eqref{eq:X2}, we lower bound the minimum distance of $h_1\msf{X}_1+h_{2}\msf{X}_{2,1}$ as follows.
\begin{align}
&d_{\min}(h_1\msf{X}_1+h_{2}\msf{X}_{2,1})\\
=&\eta \cdot d_{\min}\left(2^{\frac{\log\SNR_1}{2}}\msf{F}_{1,2}+2^{\frac{\log \SNR_1-m_{2,1}+n_{2}}{2}}\msf{F}_{2,1}\right. \nonumber \\
&\left.+2^{\frac{\log\SNR_1+n_{2}}{2}}\msf{F}_{1,1}\right) \\
=&\sqrt{\frac{1}{2^{-\log\SNR_1}\max\{E_{1}, E_{2,1}\}}} \nonumber
\end{align}

\begin{align}
&\times d_{\min}\left(\msf{F}_{1,2}+2^{\frac{n_{2}-m_{2,1}}{2}}\msf{F}_{2,1}+2^{\frac{n_{2}}{2}}\msf{F}_{1,1}\right) \\
\geq & \sqrt{\frac{6}{2^{n_1-\log\SNR_1}}} \label{eq:dmin_bound1} \\
\geq & \sqrt{3}, \label{eq:dmin_bound2}
\end{align}
where in \eqref{eq:dmin_bound1} we have used the fact that $\max\{E_{1}, E_{2,1}\} \leq \frac{2^{n_1}}{6}$ and $d_{\min}(\msf{F}_{1,2}+2^{\frac{n_{2}-m_{2,1}}{2}}\msf{F}_{2,1}+2^{\frac{n_{2}}{2}}\msf{F}_{1,1}) =1$ by applying Lemma \ref{lem:dmin} in Appendix \ref{sec:app} twice, and \eqref{eq:dmin_bound2} is due to that $n_1-\log\SNR_1 \leq 1$. From \eqref{eq:dmin_bound2}, it can be seen that the proposed constellation and power coefficient designs in \eqref{eq:X1} and \eqref{eq:X2} guarantees that the superimposed constellation $h_1\msf{X}_1+h_{2}\msf{X}_{2,1}$ has a constant minimum distance lower bound.

Note also that according to Lemma \ref{lem:dmin} in Appendix \ref{sec:app},  $\msf{F}_{1,2}+2^{\frac{n_{2}-m_{2,1}}{2}}\msf{F}_{2,1}+2^{\frac{n_{2}}{2}}\msf{F}_{1,1}$ forms a regular QAM with zero mean, minimum distance 1, and cardinality $2^{n_1}$. In this case, \eqref{eq:I_dmin1} can be refined to
\begin{align}
I(h_1\msf{X}_1+&h_{2}\msf{X}_{2,1};\msf{Y}_1)
\geq H(h_1\msf{X}_1+h_{2}\msf{X}_{2,1})-\log\left(\frac{2\pi e}{12}\right) \nonumber\\
&-\log\left(1+\frac{12}{d^2_{\min}(h_1\msf{X}_1+h_{2}\msf{X}_{2,1})}\right)\label{eq:I_bound}\\
\overset{\eqref{eq:dmin_bound2}}{=}&m_1+m_{2,1}-\log\left(\frac{5\pi e}{6}\right),\label{eq:d_bound}
\end{align}
where \eqref{eq:I_bound} is obtained by following the proof of \cite[Lemma 5]{9535131} and setting the packing random variable therein $\msf{U}\overset{\text{unif}}{\sim}\mathbb{Z}^2$ with $\boldsymbol{G}_{\mathbb{Z}^2}=d_{\min}(h_1\msf{X}_1+h_{2}\msf{X}_{2,1})\boldsymbol{I}_2$, where $\boldsymbol{I}_2$ denotes a $2 \times 2$ identity matrix. Substituting \eqref{eq:d_bound} into \eqref{eq:I1_GMAC} gives
\begin{align}\label{eq:I1_final}
I(\msf{X}_1;\msf{Y}_1) \geq m_1-\log\left(\frac{5\pi e}{6}\right).
\end{align}
Following the steps from \eqref{eq:I1_step1} to \eqref{eq:I1_final}, we obtain the mutual information for user 2 under TIN decoding as
\begin{align}\label{eq:I2_final}
I(\msf{X}_{2,t};\msf{Y}_t) \geq m_{2,t}-\log\left(\frac{5\pi e}{6}\right),
\end{align}
where $t\in\{1,2\}$ is the sub-block index. Since the deterministic rate region $(m_1,m_{2,1},m_{2,2})$ is within 1 bit to the corresponding Gaussian counterpart \cite{Avestimehr11}, the proposed scheme with TIN is capable of achieving the capacity region of the Gaussian MAC to within a \emph{constant gap} for \emph{all} channel parameters according to \eqref{eq:I1_final} and \eqref{eq:I2_final}.

According to Theorem \ref{prof:u1}, the minimum distance affects both mutual information and dispersion. Unfortunately, deriving a closed-form bound for the dispersion term based on discrete constellations is very challenging. In the next section, we will show numerically that the proposed scheme leads to lower dispersion than that under Gaussian signaling.



\section{Numerical Results}

\begin{figure}[t!]
	\centering
\includegraphics[width=\linewidth]{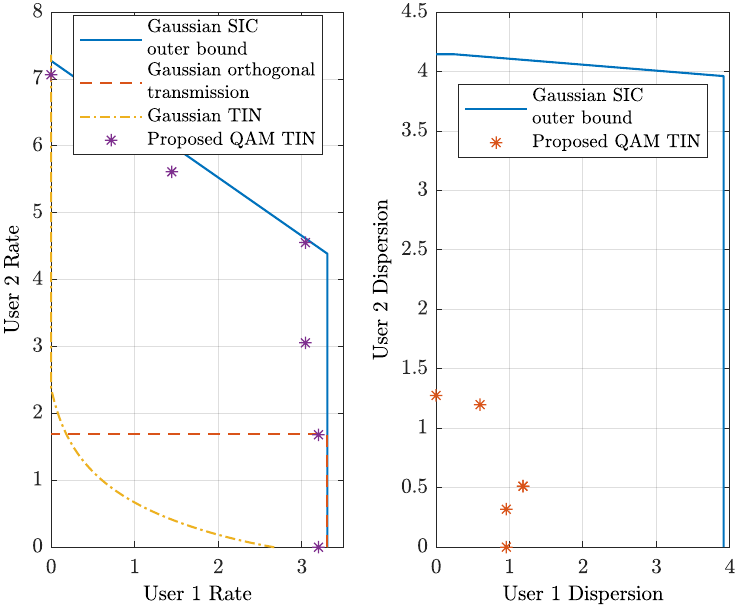}
\caption{(a) Achievable rate pairs (bits/s/Hz); (b) dispersion pairs (bits$^2$/s/Hz).}
\label{fig:rate1}
\end{figure}

In this section, we provide numerical results to show the finite blocklength achievable rates of the proposed scheme using QAM with TIN. We consider the channel model in Sec. \ref{sec:model}, where $(\SNR_1,\SNR_2)=(12,24)$ dB, $(\epsilon_1,\epsilon_2)=(10^{-6},10^{-5})$, and $(N_1,N_2)=(150,200)$. The achievable rate pairs of the proposed scheme are computed by taking the first two terms of \eqref{eq:ukrate} and \eqref{eq:u1rate} and shown in Fig. \ref{fig:rate1}(a). The proposed scheme uses the following modulation orders $(m_1,m_{2,1},m_{2,2})=(0,8,8)$, $(2,6,8),(4,4,8),(4,2,8),(4,0,8),(4,0,0)$, which corresponds to the rate pairs orders from left to right in the figure.


In the same figure, we also include the achievable rates of three benchmark schemes based on Gaussian signaling and perfect SIC, orthogonal transmission, and TIN, respectively. For the first benchmark scheme based on perfect SIC, the rate region is generated by using the convex combination of all corner points. However, it should be noted that under finite blocklength, SIC can fail and time-sharing is not able to provide a convex region \cite{6665138}. Thus, the rate region of the first benchmark scheme is an outer bound of the actual rate region under imperfect SIC and time-sharing, which is a hypothetical rate region used for comparison purposes only. It should be kept in mind that in the heterogeneous case, the messages of the more urgent user have to be decoded without waiting for the reception of the whole transmission block of the other user; therefore, SIC is infeasible in URLLC communication scenarios. The orthogonal transmission means that user 2 does not transmit any symbols for the first sub-block such that $x_2[j]=0$ for $j=1,\ldots,N_1$. In this way, the transmission of both users do not interfere with each other.


Fig. \ref{fig:rate1}(a) shows that the rate region of Gaussian TIN is the smallest. Intuitively, Gaussian interference lacks structures which may be difficult to be exploited when TIN is used. In contrast, thanks to the structural interference introduced by our carefully designed discrete input signaling, the proposed scheme with QAM and TIN can significantly outperform the second and third benchmark schemes based on Gaussian signaling. More importantly, the proposed scheme can perform very close to the benchmark scheme based on Gaussian signaling and perfect SIC.


To further investigate the impacts of the proposed design on the second-order term of the achievable rate, we plot the dispersion pairs of the proposed scheme and those of the first benchmark scheme in Fig. \ref{fig:rate1}(b). Since the second and third benchmark schemes perform poorly as shown in Fig. \ref{fig:rate1}(a), there is no need to discuss their dispersions further. Observe that the proposed scheme has much smaller dispersion pairs than the benchmark scheme with perfect SIC. It is also worth emphasizing that the main features of URLLC communications are short blocklength and ultra-low error probability. This means that even a small increase in the dispersion can have a non-negligible impact on the achievable rate. Therefore, a small dispersion is desirable. To sum up, the superior performance of the proposed scheme with QAM and TIN is owing to close-to-capacity mutual information and smaller dispersion. As a result, we believe that the proposed scheme is promising for supporting heterogeneous services in the uplink.

\section{Conclusion}
In this paper, we have investigated the problem of the coexistence of heterogeneous services in the uplink. We have constructed a capacity-achieving coding scheme with TIN for the deterministic MAC with heterogeneous interference. Then, we have translated the aforementioned coding scheme into the coded modulation scheme for the Gaussian MAC with heterogeneous interference. We have proved that the mutual information of the proposed scheme with QAM and TIN is constant gap optimal. Numerical results have shown that the proposed scheme can achieve rate pairs close to the benchmark scheme based on Gaussian signaling and perfect SIC.

\appendices
\section{A Useful Lemma for Superimposed Constellations}\label{sec:app}
\begin{lemma}\label{lem:dmin}
Consider a pair of regular QAM constellations $(\Lambda_1,\Lambda_2)$ with zero mean and $d_{\min}(\Lambda_1) = d_{\min}(\Lambda_2)=\delta>0$. The superimposed constellation $\Lambda_1+2^{\frac{\log|\Lambda_1|}{2}}\Lambda_2$ is a regular QAM with zero mean, $d_{\min}(\Lambda_1+2^{\frac{\log|\Lambda_1|}{2}}\Lambda_2) = \delta$, and cardinality $|\Lambda_1|\cdot|\Lambda_2|$.
\end{lemma}
\begin{IEEEproof}
We first look at the superposition of $\Lambda_1$ and $\Lambda_2$ in real parts by fixing the imaginary parts, i.e., $\Re(\Lambda_1)+2^{\frac{\log|\Lambda_1|}{2}}\Re(\Lambda_2)$. For $k\in\{1,2\}$, we have
\begin{align}\label{eq:PAM_set}
\Re(\Lambda_k)=\left\{-\frac{2^{\frac{\log|\Lambda_k|}{2}-1}}{2}\delta, -\frac{2^{\frac{\log|\Lambda_k|}{2}+1}}{2}\delta,\ldots, \frac{2^{\frac{\log|\Lambda_k|}{2}-1}}{2}\delta\right\}.
\end{align}
Clearly, for any two neighboring constellation points $\lambda_{k,1},\lambda_{k,2} \in \Re(\Lambda_k)$ and $\lambda_{k,1}<\lambda_{k,2}$, we have
\begin{align}\label{eq:L2_dmin}
\lambda_{k,2}-\lambda_{k,1} = \delta.
\end{align}
Next, we treat $\Re(\Lambda_1)$ as a cluster and compute the inter-cluster distance of $\Re(\Lambda_1)+2^{\frac{\log|\Lambda_1|}{2}}\Re(\Lambda_2)$ as
\begin{align}
d(\Re(\Lambda_1)&+\lambda_{2,1},\Re(\Lambda_1)+\lambda_{2,2})=-\min\{\Re(\Lambda_1)\}\nonumber \\
=&+2^{\frac{\log|\Lambda_1|}{2}}\lambda_{2,2}
-\max\{\Re(\Lambda_1)\}-2^{\frac{\log|\Lambda_1|}{2}}\lambda_{2,1} \\
=&-\frac{2^{\frac{\log|\Lambda_1|}{2}-1}}{2}\delta\cdot2+2^{\frac{\log|\Lambda_1|}{2}}\delta \label{eq:dcl_step1} \\
=& \delta, \label{eq:dcl_step2}
\end{align}
where \eqref{eq:dcl_step1} follows by using \eqref{eq:PAM_set} with $k=1$ and \eqref{eq:L2_dmin} with $k=2$.
By \eqref{eq:L2_dmin} and \eqref{eq:dcl_step2}, we know that for any pair of neighboring points $\lambda'_1,\lambda'_2\in\Re(\Lambda_1)+2^{\frac{\log|\Lambda_1|}{2}}\Re(\Lambda_2)$ and $\lambda'_1<\lambda'_2$, we have
\begin{align}
\lambda'_2-\lambda'_1=\delta.
\end{align}
As a result, the following holds.
\begin{align}\label{eq:sup_real}
\Re(\Lambda_1)&+2^{\frac{\log|\Lambda_1|}{2}}\Re(\Lambda_2) = \left\{-\frac{2^{\frac{\log(|\Lambda_1|\cdot|\Lambda_2|)}{2}-1}}{2}\delta, \right. \nonumber \\
&\left.-\frac{2^{\frac{\log(|\Lambda_1|\cdot|\Lambda_2|)}{2}+1}}{2}\delta,\ldots, \frac{2^{\frac{\log(|\Lambda_1|\cdot|\Lambda_2|)}{2}-1}}{2}\delta\right\}.
\end{align}
Due to the symmetry property of the regular QAM, we also have that
\begin{align}\label{eq:sup_imag}
\Im(\Lambda_1)+2^{\frac{\log|\Lambda_1|}{2}}\Im(\Lambda_2) =\Re(\Lambda_1)&+2^{\frac{\log|\Lambda_1|}{2}}\Re(\Lambda_2).
\end{align}
Combining \eqref{eq:sup_real} and \eqref{eq:sup_imag}, we arrive at the conclusion stated in Lemma \ref{lem:dmin}.
\end{IEEEproof}

\bibliographystyle{myIEEEtran.bst}
\bibliography{MinQiu}

\begin{thebibliography}{10}
\providecommand{\url}[1]{#1}
\csname url@samestyle\endcsname
\providecommand{\newblock}{\relax}
\providecommand{\bibinfo}[2]{#2}
\providecommand{\BIBentrySTDinterwordspacing}{\spaceskip=0pt\relax}
\providecommand{\BIBentryALTinterwordstretchfactor}{4}
\providecommand{\BIBentryALTinterwordspacing}{\spaceskip=\fontdimen2\font plus
\BIBentryALTinterwordstretchfactor\fontdimen3\font minus
  \fontdimen4\font\relax}
\providecommand{\BIBforeignlanguage}[2]{{%
\expandafter\ifx\csname l@#1\endcsname\relax
\typeout{** WARNING: IEEEtran.bst: No hyphenation pattern has been}%
\typeout{** loaded for the language `#1'. Using the pattern for}%
\typeout{** the default language instead.}%
\else
\language=\csname l@#1\endcsname
\fi
#2}}
\providecommand{\BIBdecl}{\relax}
\BIBdecl

\bibitem{8004168}
H.~Zhang, N.~Liu, X.~Chu, K.~Long, A.-H. Aghvami, and V.~C.~M. Leung, ``Network
  slicing based 5{G} and future mobile networks: Mobility, resource management,
  and challenges,'' \emph{IEEE Commun. Mag.}, vol.~55, no.~8, pp. 138--145,
  2017.

\bibitem{8476595}
P.~Popovski, K.~F. Trillingsgaard, O.~Simeone, and G.~Durisi, ``5{G} wireless
  network slicing for e{MBB}, {URLLC}, and m{MTC}: A communication-theoretic
  view,'' \emph{IEEE Access}, vol.~6, pp. 55\,765--55\,779, 2018.

\bibitem{8647460}
R.~Kassab, O.~Simeone, and P.~Popovski, ``Coexistence of {URLLC} and e{MBB}
  services in the {C-RAN} uplink: An information-theoretic study,'' in
  \emph{Proc. IEEE Globecom}, 2018, pp. 1--6.

\bibitem{9097306}
M.~B. Shahab, R.~Abbas, M.~Shirvanimoghaddam, and S.~J. Johnson, ``Grant-free
  non-orthogonal multiple access for {IoT}: A survey,'' \emph{IEEE Commun.
  Surveys Tut.}, vol.~22, no.~3, pp. 1805--1838, 2020.

\bibitem{9562192}
O.~Dizdar, Y.~Mao, Y.~Xu, P.~Zhu, and B.~Clerckx, ``Rate-splitting multiple
  access for enhanced {URLLC} and e{MBB} in 6{G},'' in \emph{Proc. 17th Int.
  Symp. Wireless Commun. Syst. (ISWCS)}, Sep. 2021, pp. 1--6.

\bibitem{9838392}
P.-H. Lin, S.-C. Lin, P.-W. Chen, M.~Mross, and E.~A. Jorswieck, ``Rate region
  of {G}aussian broadcast channels with heterogeneous blocklength
  constraints,'' in \emph{Proc. IEEE Int. Conf. Commun. (ICC)}, May 2022, pp.
  2144--2150.

\bibitem{Cover:2006:EIT:1146355}
T.~M. Cover and J.~A. Thomas, \emph{Elements of Information Theory}.\hskip 1em
  plus 0.5em minus 0.4em\relax New York, NY, USA: Wiley-Interscience, 2006.

\bibitem{5452208}
Y.~Polyanskiy, H.~V. Poor, and S.~Verdu, ``Channel coding rate in the finite
  blocklength regime,'' \emph{IEEE Trans. Inf. Theory}, vol.~56, no.~5, pp.
  2307--2359, May 2010.

\bibitem{8345745}
X.~Sun, S.~Yan, N.~Yang, Z.~Ding, C.~Shen, and Z.~Zhong, ``Short-packet
  downlink transmission with non-orthogonal multiple access,'' \emph{IEEE
  Trans. Wireless Commun.}, vol.~17, no.~7, pp. 4550--4564, Jul. 2018.

\bibitem{7300429}
E.~MolavianJazi and J.~N. Laneman, ``A second-order achievable rate region for
  {G}aussian multi-access channels via a central limit theorem for functions,''
  \emph{IEEE Trans. Inf. Theory}, vol.~61, no.~12, pp. 6719--6733, 2015.

\bibitem{9535162}
R.~C. Yavas, V.~Kostina, and M.~Effros, ``Gaussian multiple and random access
  channels: Finite-blocklength analysis,'' \emph{IEEE Trans. Inf. Theory},
  vol.~67, no.~11, pp. 6983--7009, 2021.

\bibitem{TS138212_v16p8}
3GPP, ``5{G};{NR}; {M}ultiplexing and channel coding,'' {3rd Generation
  Partnership Project (3GPP)}, TR 38.212, Jan. 2022.

\bibitem{8291591}
M.~Qiu, Y.-C. Huang, S.-L. Shieh, and J.~Yuan, ``A lattice-partition framework
  of downlink non-orthogonal multiple access without {SIC},'' \emph{IEEE Trans.
  Commun.}, vol.~66, no.~6, pp. 2532 -- 2546, Jun. 2018.

\bibitem{9535131}
M.~Qiu, Y.-C. Huang, and J.~Yuan, ``Discrete signaling and treating
  interference as noise for the {G}aussian interference channel,'' \emph{IEEE
  Trans. Inf. Theory}, vol.~67, no.~11, pp. 7253--7284, Nov. 2021.

\bibitem{JSACQiu22}
M.~Qiu, Y.-C. Huang, and J.~Yuan, ``Downlink transmission with heterogeneous
  {URLLC} services: Discrete signaling with single-user decoding,'' \emph{IEEE
  J. Sel. Areas Commun.}, vol.~41, no.~7, pp. 2261--2277, 2023.

\bibitem{Avestimehr11}
A.~S. Avestimehr, S.~N. Diggavi, and D.~N.~C. Tse, ``Wireless network
  information flow: A deterministic approach,'' \emph{IEEE Trans. Inf. Theory},
  vol.~57, no.~4, pp. 1872--1905, 2011.

\bibitem{CIT-019}
A.~{Guillén i Fàbregas}, A.~Martinez, and G.~Caire, ``Bit-interleaved coded
  modulation,'' \emph{Found. Trends Commun. Inf. Theory}, vol.~5, no. 1–2,
  pp. 1--153, 2008.

\bibitem{6665138}
V.~Y.~F. Tan and O.~Kosut, ``On the dispersions of three network information
  theory problems,'' \emph{IEEE Trans. Inf. Theory}, vol.~60, no.~2, pp.
  881--903, 2014.

\end{thebibliography}

\end{document}